# Off-axis electron holography and microstructure of $Ba_{0.5}Sr_{0.5}TiO_3$ thin film grown on $LaAlO_3$

H. F. Tian[1], H. C. Yu[1], X. H. Zhu[2], Y. G. Wang[1], D. N. Zheng[2], H. X.Yang[1] and J. Q. Li[*]

[1]Beijing Laboratory of Electron Microscopy, Institute of Physics, Chinese Academy of Science, Beijing, 100080, China

[2]National Laboratory for Superconductivity, Institute of Physics, Chinese Academy of Sciences, Beijing, 100080, China

Epitaxial $Ba_{0.5}Sr_{0.5}TiO_3$ thin films grown on the (001) $LaAlO_3$ substrates with the ferroelectric transition of about 250K have been investigated by TEM and off-axis electron holography. Cross-sectional TEM observations show that the 350nm-thick $Ba_{0.5}Sr_{0.5}TiO_3$ film has a sharp interface with notable misfit dislocations. Off-axis electron holographic measurements reveal that, at low temperatures, the ferroelectric polarization results in systematic accumulations of negative charges on the interface and positive charges on the film surface, and, at room temperature, certain charges could only accumulate at the interfacial dislocations and other defective areas.

[*]Corresponding author. Fax: +86-10-8264-9531/0215.

*Email address:* ljq@aphy.iphy.ac.cn

Ferroelectric $(Ba_{1-x}Sr_x)TiO_3$ thin films have recently become very attractive due to their large dielectric constant ($\varepsilon_r$) and high tunability ($\varepsilon_r$); the dielectric constant of this material can vary from a few hundreds to thousands by adjusting the Ba/Sr ratio, grain size, and temperature. Its Curie temperature decreases almost linearly with Sr content [1], and dielectric constant exhibits remarkable tunability under an applied dc electric field near the Curie temperature [2-6]. These distinctive electric properties imply the potential applications of $(Ba_{1-x}Sr_x)TiO_3$ thin films in the high density dynamic random access memories, smart card memories, and tunable microwave devices [2, 3, 7-10]. The epitaxial $Ba_{1-x}Sr_xTiO_3$ films deposited on $LaAlO_3$ and MgO substrates by pulsed laser deposition have been suggested to be good candidates for various tunable microwave devices. High quality films which can minimize the loss of the tangent are requested for fabrication of the reliable and high performance devices. Because of the microstructure-sensitivity of the $Ba_{1-x}Sr_xTiO_3$ films, detailed investigations of the microstructure and correlation to the physical behaviors of the products are highly desirable for synthesis method evaluation and further application. In this paper, we reported on the microstructure measured by cross-section TEM and space-charge accumulations on defects, interface and surface in $Ba_{0.5}Sr_{0.5}TiO_3/LaAlO_3$ (BST/LAO) identified by off-axis electron holography.

The BST thin films on (001) LAO substrate studied in present work were prepared by pulsed laser deposition. Firstly, the BST films were grown under an oxygen pressure of about 25 Pa and at the temperature of 800°C to a thickness of about 200nm in ten minutes, and then the films were post-annealed at 500°C under one atmosphere oxygen pressure for 30 minutes. Detailed depictions on the film growth and physical properties of the as grown BST films were reported in ref.9. Electron microscopy and off-axis electron holography observations were carried out utilizing a Philips CM200/FEG transmission electron microscope equipped with the electron biprism located in the selected area aperture position. Samples for cross-section TEM and electron holography investigations were prepared using a standard procedure consisting of gluing, cutting, mechanical polishing, dimpling, and ion milling. The holograms were acquired with a Gatan 794 multiscan CCD camera, and processed using Gatan Digital Micrograph software including "Holowork" package. Off-axis electron holography provides information about the phase and amplitude of the electron wave after traversing through a specimen. This technique has been frequently applied to study ferroelectric materials [11-13] and was proved as one of effective techniques in determining the characteristics of ferroelectric materials.

Experimental results from either TEM observations or x-ray diffraction (not shown) indicate that the $Ba_{0.5}Sr_{0.5}TiO_3$ film is epitaxially grown along the <001> direction on the (001) $LaAlO_3$ substrate. Fig. 1a is a bright-field TEM image displaying the morphology of the as-grown film with a uniform thickness of about 350nm. The interface without reaction products between the film and substrate is clean and sharp as confirmed by high-resolution TEM image shown in Fig.1b. A remarkable feature of the film is the columnar structure with V-shaped grain boundaries as indicated with the solid lines; this structural feature is considered to arise from small relative rotations between the crystalline grains [10]. The presence of numerous steps and terraces on film surface indicates different growing rates of the subgrains during film preparation. The well-defined epitaxial relationship between the film and substrate is illustrated by the corresponding electron diffraction pattern inserted in Fig.1a. Two sets of electron diffraction spots, arising respectively from the BST film and the LAO substrate, can be unambiguously indexed based on $LaAlO_3$ structure (a cubic cell with the lattice parameter of 0.378nm) and $Ba_{0.5}Sr_{0.5}TiO_3$ structure with the lattice constant of 0.394nm. This pattern clearly exhibits the orientation relationship of $[010]_{BST}//[010]_{LAO}$ and $[001]_{BST}//[001]_{LAO}$. A lattice mismatch of ~4.2% between $Ba_{0.5}Sr_{0.5}TiO_3$ and $LaAlO_3$ can be obtained from splits of the diffraction spots. This lattice mismatch could result in distinctive dislocations on the interfacial area as discussed in following context. Fig.1b shows a high-resolution TEM image of the interface structure emphasizing the perfect quality of epitaxial growth and the sharp interface between the film and substrate. Fig.1c displays a high resolution TEM image showing the presence of numerous misfit dislocations on the interface. This dislocation array with a pseudo-periodicity of around L=7.5nm is roughly coincident with theoretical prediction ($L_d=a_1a_2/(a_1-a_2)$= 19×0.38nm=7.2nm). Fig.1d schematically illustrates an atomic structural model with the characteristic of an edge-dislocation along the interface.

In order to determine the charge distribution and polarization of the BST film, we have performed a series of off-axis electron holography measurements at low temperatures. When the electron beams pass through a local electric field, the electrostatic potential induces a local differential phase shift in the exit electron wave function. The phase shift directly proportional to the electrostatic potential distribution [14] can be recorded in the hologram formed by interference of the perturbed exit electron wave function with the reference wave in vacuum. Interaction between the incident electron wave and the electrostatic potential within specimen can be recorded with high spatial resolution in terms of electron hologram, which provides information about the phase of the exit electron wave function via reconstruction of the interference pattern.

Below the Curie temperature of 250K, the $Ba_{0.5}Sr_{0.5}TiO_3$ film transforms into ferroelectric state, which results in the charge redistributions due to ferroelectric polarization. In order to investigate the possible ferroelectric ordering, ferroelectric domain walls or the spontaneous polarization, the charge distribution in the film has been measured at the temperature of 120K. Fig.2a is a hologram taken at the interface region without edge dislocations under a positive bias potential of 90 Volt. Because of the limited interference scope in our experiment, the hologram containing both the vacuum and the sample edge was taken separately for correction of the phase profile. A reference hologram without the sample was acquired for correction of holographic fringe distortions due to the spurious effects caused by the image system, the inhomgeneities contamination in the biprism wire, and the nonisoplanicity of the incident illumination [15]. The Hanning window with applying 1 order was used in reconstruction to reduce the edge effect and in compensation for the phase shift artificially produced by the Fresenl fringe from the filament electrode [16]. The reconstructed images with a size of 256×256 pixels were obtained using the primary sideband and followed by a complex division using the reference hologram. Fig.2b is the reconstructed phase image corresponding to the boxed area in Fig.2a. An averaged line scan (Fig.2c) from the substrate to the film in the reconstructed phase image shows a valley of the phase profile at the interface position, which indicates net negative charges located at the interface resulting in delay of the wave front of the exit electron wave function.

The phase change at the surface of film was also measured by off-axis electron holography. Fig.3a shows a hologram taken at the surface of the film. After reconstruction the averaged phase profile across the surface is presented in Fig.3b, in which a peak on the profile appears at the surface position, this fact directly suggests that certain positive charges are accumulated on the film surface. In combination with the experimental data obtained from the interface area, we can conclude that a spontaneous polarization occurs along the [001] axis direction in the ferroelectric state due to the movements of Ba (Sr) and O atoms in opposite directions. As a result, the net charges with opposite signs accumulated respectively on the BST/LAO interface and BST surface as detected experimentally.

The misfit dislocations with unpaired dangling bonds could capture charges to form the localized electrostatic field along the dislocation lines and therefore influence the dielectric properties of the film. Actually, room-temperature measurements of dielectric properties indicate that the $Ba_{0.5}Sr_{0.5}TiO_3$ films are in paraelectric state and has the relative dielectric constant of $\varepsilon_r\sim1200$, the loss tangent of 0.016 and the tunability of about 60%. On the other hand, careful analyses suggest that that this kind of films in general shows up a faintly ferroelectric property. Fig.4a shows a papilionaceous-shaped curve obtained under 1MHz at 300K. The presence of hysteresis in the ε-v curve demonstrates the existence of local ferroelectric orders or interfacial space charges [5] in the $Ba_{0.5}Sr_{0.5}TiO_3$ film. Fig. 4b shows a hologram taken from an area with clear interfacial dislocations by a positive bias of 120 Volt applied to the biprism. The lattice fringes superimposed on the fine holographic fringes show a typical interfacial dislocation (indicated by arrows). The reconstructed images were obtained using the primary sideband as mentioned in above context. Although there is no detectable phase change at the interface position as expected for the paraelectric state, a line scan across the misfit dislocation shows remarkable phase alternations. In order to improve the poor statistic of a single phase shift scan, averaged phase profile over 50 line scans was taken in our analysis shows obviously a positive phase jump on the dislocation center. Differentiation of the phase profile as shown in the Fig.4c could predict the charge distribution around the dislocation, which demonstrates positive charges accumulation on the dislocation core with negative charges around. In accordance with the structural model (Figure 1d) and our experimental results, this kind of charge distribution can be qualitatively understood as following: the considerable stress on the misfit dislocations could either result in local oxygen vacancies or yields outward movements of oxygen ions so that net positive charges appear locally on the center area. In addition, our measurements also show that a small amount of charges also accumulate on the intragrain boundaries at room temperature. These experimental results are consistent with the data of the dielectric measurements; the local charges and the related polarizations make ε-v curve papilionaceously shaped and give rise to the weak ferroelectric property in the room-temperature paraelectric state.

In conclusion, the $Ba_{0.5}Sr_{0.5}TiO_3$ thin films with ferroelectric transition at about 250K, grown on the (001) $LaAlO_3$ substrates by pulsed laser deposition, have a well-defined epitaxial structure with a sharp interface and notable misfit dislocations. This kind of ferroelectric films has an evident ferroelectric polarization along <001> direction at around 120K as demonstrated by off-axis electron holographic investigations. As a result, negative and positive charges are accumulated on the interface and the film surface, respectively. Room-temperature measurements of electron holography reveal local positive charge accumulation on interfacial dislocations and some other grain boundaries in the $Ba_{0.5}Sr_{0.5}TiO_3$ thin film, which could result in weak ferroelectric properties as confirmed by the dielectric data.

**Acknowledgments**

We would like to thank Prof. Y.Q. Zhou, Prof. X.F. Duan and Miss Y. Li for their assistance in preparing samples. The work reported here was supported by National Natural Science Foundation of China.

**Figure Captions**

**Figure 1:** (a) Bright-filed TEM image showing morphology of the epitaxial BST film on LAO substrate, the insert shows an electron diffraction pattern from the interface area. (b) HRTEM image showing the details of the interface of epitaxial growth of BST film. (c) HRTEM image showing the misfit dislocations on the interface. (d) Atomic structural model for an interfacial dislocation.

**Figure 2:** (a) Hologram taken at 120K for the interface area. (b) Reconstructed phase image from the hologram. (c) Phase profile showing a valley across the BST/LAO interface.

**Figure 3:** (a) Hologram taken at 120K for the surface of BST film. (b) Phase profile showing up a peak at the position of film surface.

**Figure 4:** (a) Dielectric constant as a function of applied dc bias voltage under 1MHz at 300K. (b) Electron hologram from an interface area with a notable misfit dislocation at 300K. (c) Charge distribution cross this dislocation.

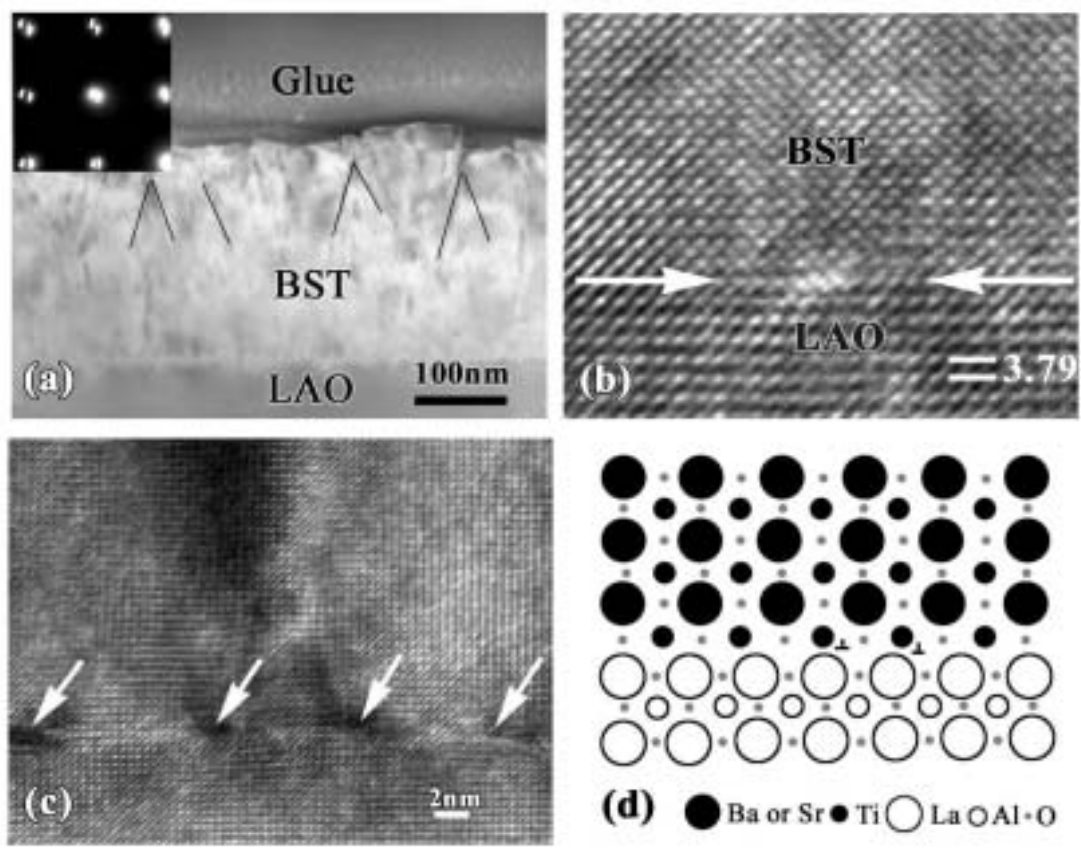

**Fig.1**

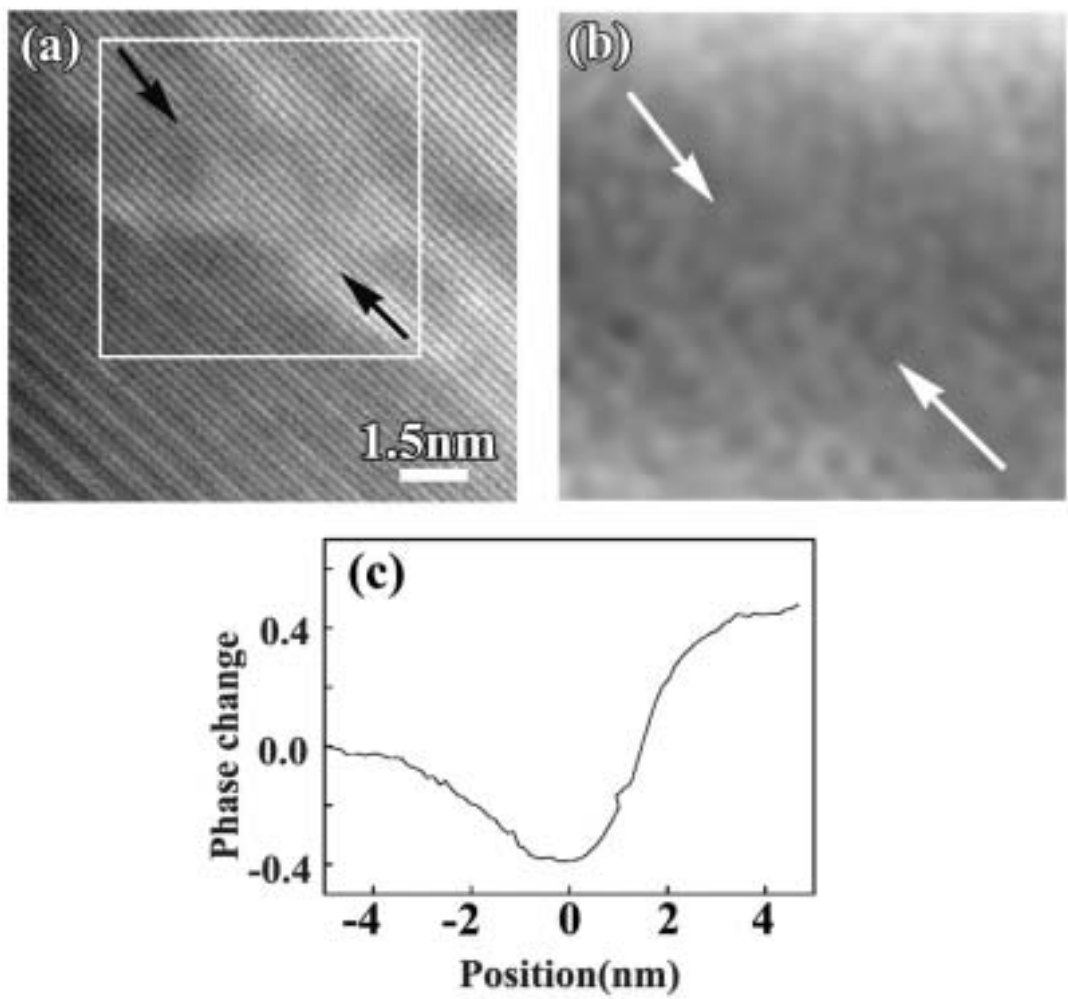

**Fig.2**

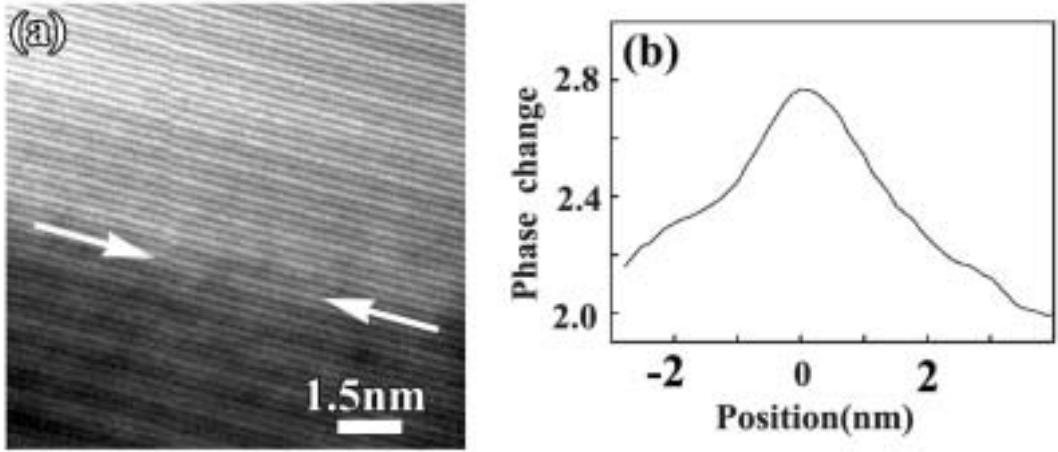

**Fig.3**

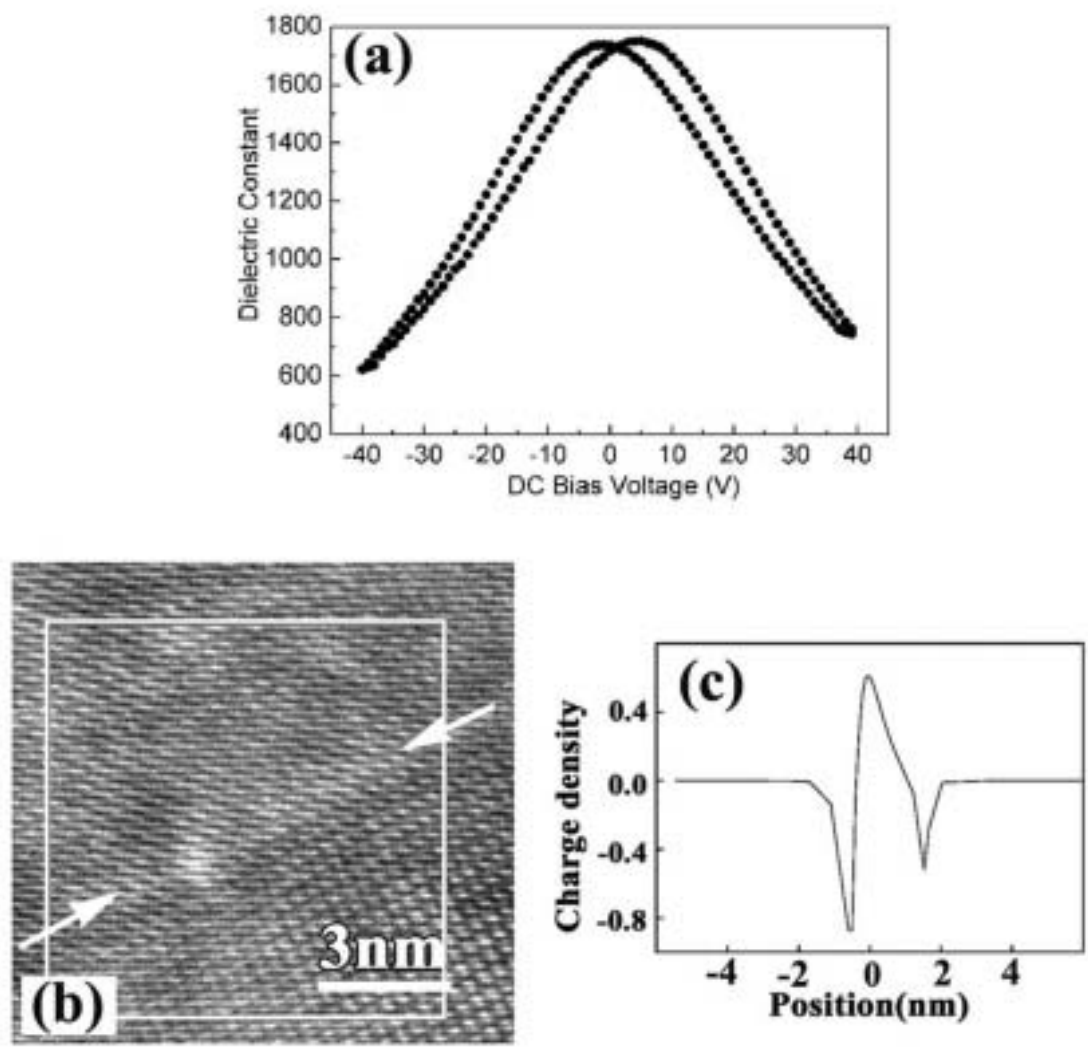

**Fig.4**